\documentclass{article}
\usepackage{amsmath}
\usepackage{dsfont}
\usepackage{graphicx}
\usepackage{xcolor}

\DeclareMathOperator*{\argmin}{arg\,min}

\title{Learning to regularize with a variational autoencoder for hydrologic inverse analysis}
\author{Daniel O'Malley$^{1,2}$, John K. Golden$^1$, and Velimir V. Vesselinov$^1$\\
{\footnotesize$^1$Computational Earth Science, Los Alamos National Laboratory, Los Alamos, NM 87545}\\
{\footnotesize$^2$Department of Computer Science and Electrical Engineering, University of Maryland,}\\{\footnotesize Baltimore County, MD 21250}}

\begin{document}

\maketitle

\begin{abstract}
Inverse problems often involve matching observational data using a physical model that takes a large number of parameters as input.
These problems tend to be under-constrained and require regularization to impose additional structure on the solution in parameter space.
A central difficulty in regularization is turning a complex conceptual model of this additional structure into a functional mathematical form to be used in the inverse analysis.
In this work we propose a method of regularization involving a machine learning technique known as a variational autoencoder (VAE).
The VAE is trained to map a low-dimensional set of latent variables with a simple structure to the high-dimensional parameter space that has a complex structure.
We train a VAE on unconditioned realizations of the parameters for a hydrological inverse problem.
These unconditioned realizations neither rely on the observational data used to perform the inverse analysis nor require any forward runs of the physical model, thus making the computational cost of generating the training data minimal.
The central benefit of this approach is that regularization is then performed on the latent variables from the VAE, which can be regularized simply.
A second benefit of this approach is that the VAE reduces the number of variables in the optimization problem, thus making gradient-based optimization more computationally efficient when adjoint methods are unavailable.
After performing regularization and optimization on the latent variables, the VAE then decodes the problem back to the original parameter space.
Our approach constitutes a novel framework for regularization and optimization, readily applicable to a wide range of inverse problems.
We call the approach RegAE.

\end{abstract}

\section{Introduction}

Assimilating observational data into computational physics models is often accomplished by solving an inverse problem.
Inverse problems essentially seek to reverse engineer an underlying model by matching to observational data, and therefore play a critical role in a variety of science and engineering fields.
In this paper we focus on the solution of an inverse problem related to subsurface hydrology where the goal of the inverse model is to infer the spatially heterogeneous hydraulic conductivity of an aquifer from observations of pressure and the hydraulic conductivity at a sparse collection of points within the flow domain.
Inverse problems are often inherently difficult, requiring computationally intensive optimization procedures to produce a solution.
Another challenge is that inverse problems tend to be under-constrained, or ill-posed, resulting in many viable solutions.
That is, many choices of the model parameters result in predictions that are in agreement with the observations.

Regularization is a technique often used to transform the inverse problem from ill-posed to well-posed by adding a term (which we call the regularization term) to the objective function that the inverse analysis aims to minimize.
Specifically, regularization seeks to impose additional desired features on the solution such as smoothness.
With regularization added to the objective function, the inverse analysis therefore tries to minimize the sum of the misfit to the observational data (which depends on the observations and forward model predictions) and the regularization term (which depends on the model parameters, but not the observations or model predictions).

One commonly used regularization technique is Tikhonov regularization \cite{tikhonov1963solution,franklin1974tikhonov}.
When the model is parameterized by one or more spatially heterogeneous fields, regularization often seeks to smooth out these fields, e.g., by encouraging the gradient (or higher derivatives) of the field to be small \cite{rudin1992nonlinear,bredies2010total}.
In the hydrogeologic context that we consider, the geostatistical approach \cite{kitanidis1983geostatistical} to inverse analysis is frequently used, and this approach relies on a regularization term based on the Mahalanobis distance \cite{mahalanobis1936generalized}.
The covariance matrix in the Mahalanobis distance effectively encodes a conceptual model of the structure of the parameters.
What we refer to as the conceptual model of the structure of the parameters is analogous to a prior in a Bayesian approach.
One can think of the conceptual model as capturing trends, correlations, or other structures in the parameters.
When this conceptual model is sufficiently complex, it can be difficult to construct a regularization term that is appropriate for use in inverse analysis.

In this paper we propose a novel approach to assist in both the regularization and optimization by utilizing a variational autoencoder (VAE) \cite{kingma2013auto}.
A VAE is a machine learning methodology that transforms a high-dimensional object into a set of low-dimensional latent variables with a simple structure and back again.
The high-dimensional object usually has some implicit structure such as an image of a celebrity face \cite{hou2017deep} or a heterogeneous subsurface parameter field (as is explored here).
This process of mapping from the high-dimensional object to the latent variables is called ``encoding''.
Similarly, there is a ``decoding'' process that maps from the latent variables back to the high-dimensional object, and the encoding and decoding processes are ideally the inverse of one another (i.e, the output of the VAE is the same as the input).

The major innovation of the methodology proposed here is an automated and straightforward way of regularizing with a VAE.
Rather than having to discover a regularization term that is suitable for the inverse problem being posed, we train a VAE with unconditioned realizations of the parameter fields that are consistent with the conceptual model of parameter structure.
The resulting latent variables of the trained VAE then create an approximately normally distributed parameter space that conforms to the existing conceptual model when mapped through the decoder.
An additional benefit of this approach is that the optimization necessary to fit our model to observational data can be performed in this lower-dimensional space and is therefore more computationally efficient.
The regularized, optimized answer in latent space can then be decoded back in to our original parameter space.

This approach to regularization is a form of unsupervised machine learning.
We contrast this with supervised machine learning which might, e.g., be used to produce a fast, approximate version of the forward model (i.e., a surrogate model).
One of the challenges with the supervised machine learning approach to develop surrogate models is that the number of model runs required to train the surrogate model is often very large when the forward model is complex.
In order for this process to produce significant computational savings, the number of runs of the surrogate model must be even larger.
For example, if the surrogate model is to be used in an inverse analysis, the number of forward model runs required to train the surrogate model should be less than the number of forward model runs required to perform the inverse analysis.
Our approach requires no runs of the forward model and therefore avoids this problem.
We call our approach RegAE since it does \emph{Reg}ularization with a variational \emph{A}uto\emph{E}ncoder.

The remainder of this manuscript is organized as follows.
In section \ref{sec:methods}, we describe the methods that are used to realize RegAE.
In section \ref{sec:results}, we describe the results of applying RegAE to solve a hydrologic inverse problem.
In section \ref{sec:discussion}, we discuss possible improvements to the methods that are utilized here.
Finally, we present our conclusions in section \ref{sec:conclusion}.

\section{Methods}\label{sec:methods}
\begin{figure}
	\includegraphics[width=\textwidth]{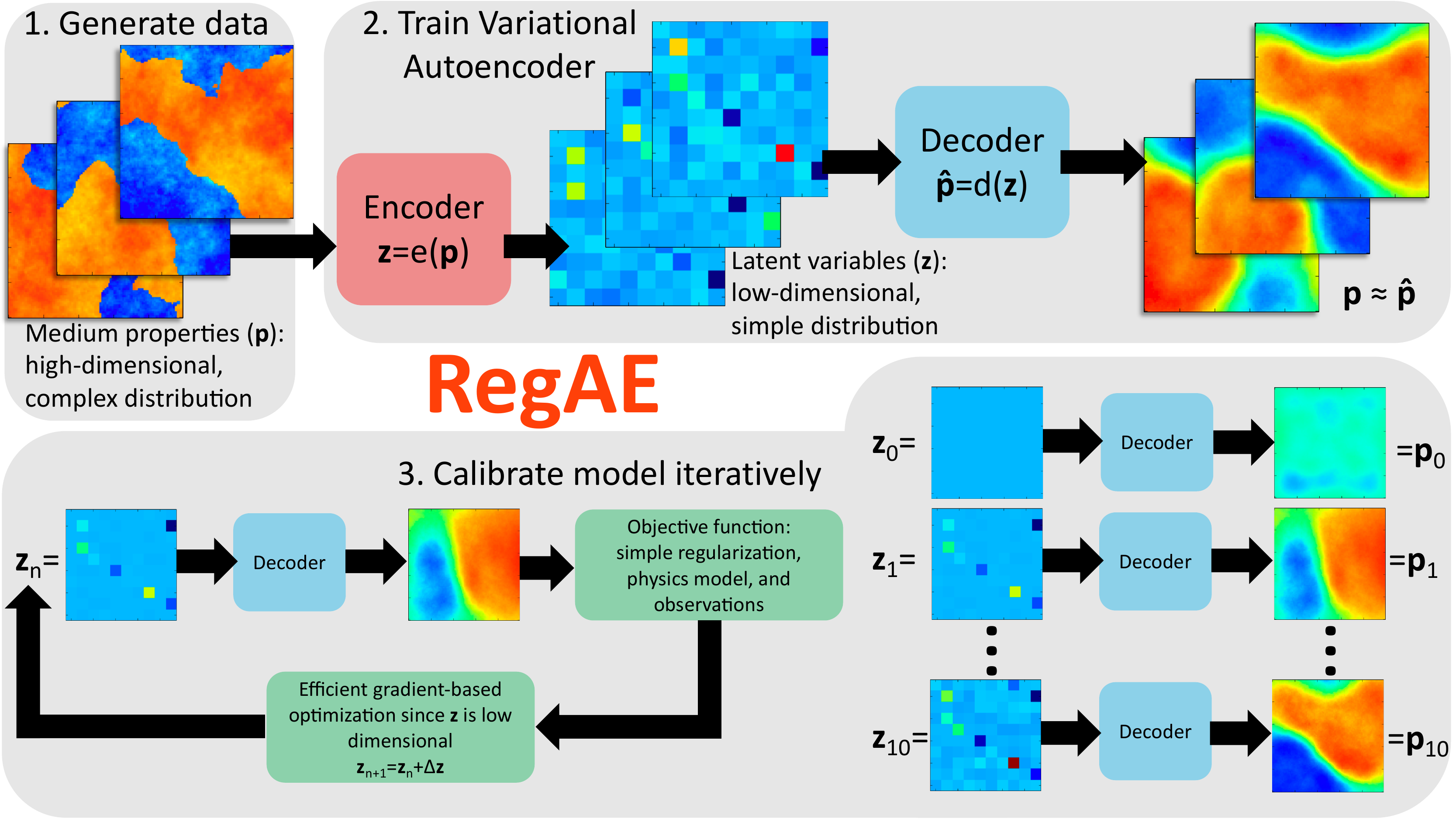}
	\caption{The RegAE inverse method leverages a VAE for dimensionality reduction and an automated approach to regularization.
	Calibration is performed on the low-dimensional latent variables ($\mathbf{z}$) rather than the high-dimensional medium properties, making gradient computations efficient.
	The simple distribution of the latent variables makes regularization simple.}
	\label{fig:workflow}
\end{figure}

Our method is composed of three steps:
\begin{enumerate}
	\item Generate unconditioned random realizations of the parameter field.
	\item Train the VAE.
	\item Use the VAE combined with a physical model to perform the inverse analysis using gradient-based optimization methods.
\end{enumerate}
We will use the notation $\mathbf{p}$ or $\mathbf{p}^i$ to denote a physical parameter field in vector form, $\mathbf{z}$ to denote the latent variables from the VAE, $e(\cdot)$ to denote the encoder, $d(\cdot)$ to denote the decoder, $h(\cdot)$ to denote the forward model, and $\hat{\textbf{h}}$ to denote a vector of observations that inform the inverse analysis.
We also let $n_p$, $n_z$, and $n_h$ be the number of components of $\mathbf{p}$, $\mathbf{z}$, and $\hat{\mathbf{h}}$, respectively.
Note that one of the implied concepts here is that the number of latent variables is much less than the number of physical parameters, i.e., $n_z\ll n_p$.
The forward model we employ here simulates single phase flow in heterogeneous porous media.
The parameters, $\mathbf{p}$, describe the hydraulic conductivity of the porous medium and the observations, $\hat{\mathbf{h}}$, are of the fluid pressure and hydraulic conductivity at various locations throughout the domain.
Figure \ref{fig:workflow} illustrates the RegAE workflow.

\subsection{Data Generation}
The first step in our approach to inverse analysis is to generate a sequence of realizations, $\mathbf{p}_i$ for $i=1,2,\ldots,N$.
These realizations should be independent of the observational data, $\hat{\mathbf{h}}$, that will be used for calibration by the inverse analysis.
In general, these realizations could come from a variety of sources such as experiments, statistical models, or physical simulations that produce a realization of the parameter field as output.
The main caveat is that a sufficient number of realizations must be generated so as to effectively train the VAE.
Realizations of the parameter fields that come from statistical models or physical simulations are a natural fit here because this data generation process can be automated by a computer in an embarrassingly parallel fashion.

In this work, we utilize a statistical model of $\mathbf{p}$ that results in the hydraulic conductivity field having two hydrogeologic facies that have distinct properties.
In the course of generating each hydraulic conductivity realization, three fields are generated all of which have a fractal character \cite{peitgen1988science}.
The code for generating the fields is part of the Fast Fourier Transform Random Field (FFTRF) module of the GeostatInversion.jl \cite{geostatinversion2019} package for the Julia \cite{bezanson2017julia} programming language.
Each of these fields is characterized by a mean, variance, and a parameter from the power-law spectrum which we call $\beta$.
These fractal fields do not have a correlation length, but $\beta$ plays a role similar to the correlation length.
As $\beta$ decreases the fields become smoother (analogous to increasing the correlation length), and as $\beta$ increases the fields become rougher (analogous to decreasing the correlation length).
Two of the fields (``Conductivity 1'' and ``Conductivity 2'') are used to define the hydraulic conductivity within each of the hydrogeologic facies.
The third field (``Split'') is used to determine which of the two hydrogeologic facies is present at a given location.
The parameters used for each of these fields is given in Table \ref{tab:fieldparams}.

\begin{table}[]
	\begin{tabular}{|l|l|l|l|}
		\hline
		Field Name & Mean $(L/T)$ & Variance $(L^2/T^2)$ & $\beta$ $(-)$ \\
		\hline
		Conductivity 1 & $10^{-5}$ & $\frac{1}{2}$ & -3 \\
		Conductivity 2 & $10^{-8}$ & $\frac{1}{2}$ & $-\frac{5}{2}$ \\
		Split & $10^{-8}$ & 1 & $-\frac{9}{2}$ \\
		\hline
	\end{tabular}
	\caption{The statistical parameters used to generate the hydraulic conductivity data.}
	\label{tab:fieldparams}
\end{table}

These three fields are combined as follows to produce a hydraulic conductivity field, $\mathbf{p}$.
First a random number, $q$, is drawn uniformly between $1/4$ and $3/4$.
At each point in space, if the value of the ``Split'' field is below the $q^{\mathrm{th}}$ quantile, then the value of the hydraulic conductivity is taken from the ``Conductivity 1'' field and is taken from the ``Conductivity 2'' otherwise.
This effectively means that a fraction $q$ of the domain will be covered by the ``Conductivity 1'' field and a fraction $1-q$ will be covered by the ``Conductivity 2'' field.

We explored datasets with $N=10^4$, $N=10^5$, and $N=10^6$ realizations.
We found some improvement increasing from $N=10^4$ to $N=10^5$, but increasing to $N=10^6$ did not provide significant further improvements.
The number of realizations that are needed to train the VAE will vary depending on the complexity of both the distribution of $\mathbf{p}$ and of the VAE.
Generally, the more complex either is, the larger $N$ will need to be.

\subsection{Variational Autoencoder}

VAEs \cite{kingma2013auto} are generative machine learning models built on top of neural networks and trained via unsupervised learning.
They have found widespread application, generating realistic faces, handwriting, and music (see \cite{doersch2016} for a review and introduction).
As with other types of autoencoders, such as sparse and denoising, VAEs are composed of an encoder network $e(\cdot)$ and a decoder network $d(\cdot)$.
The encoder network reduces the dimension, mapping a high-dimensional space (such as pixels in an image) on to a smaller space.
This smaller space is often referred to as latent space, which is composed of so-called latent variables.
For example, in the context of an autoencoder trained on handwriting images, latent variables could include not only the underlying character, but also such properties as stroke width and angle (therefore retaining more data than other types of networks designed purely for classification purposes).
The decoder network works in reverse, taking a collection of latent variables and mapping them back to the original higher-dimension space.
The encoder and decoder networks are trained in tandem such that $d(e(\cdot)) \approx \text{id}(\cdot)$.

One of the unique feature of VAEs, when compared to other autoencoders, is that the output of the encoder describes a probability distribution of the latent variables, rather than concrete values.
That is, for a given input $X$ the VAE produces a collection of means $\mu(X)$ and standard deviations $\sigma(X)$.
The central benefit of this approach is that it creates a locally continuous latent space.
This means that small changes in the latent variables result in small changes in the decoded values.
The locally continous nature of the latent space is critical for ensuring that the gradients used in the inverse analysis are sensible.

However, while similar items will map to some locally continuous portion of latent space, there could still be large untrained portions of training space separating the locally continuous regions.
For example, returning to the case of handwriting recognition, images of the character ``1'' will form a smooth region in encoded space, but there is no guarantee that they will be anywhere near the space associated with images of ``2''.
VAEs are constructed in such a way as to avoid this problem, thus creating a globally continuous latent space approximately following a normal distribution with mean 0 and covariance matrix being the identity.
This is done by minimizing the Kullback-Leibler divergence \cite{kullback1997information} between the distribution of the latent variables during training and a $N(0, I)$ distribution.
To be more specific, the loss function used in training includes a term for the error in the reconstruction after the encoding and decoding process as well as a term for the Kullback-Leibler divergence.
The globally continuous nature of the latent space in a VAE is critical for the inverse analysis, because the optimization algorithm must move through the latent space continuously to (ideally) find the global minimum.

In the context of inverse analysis and regularization, VAEs thus addresses two significant issues.
First, there are far fewer latent variables than input variables, thus significantly reducing computational overhead for calculating gradients.
Second, the space of latent variables follows a normal distribution, thus dramatically simplifying the regularization process.
The VAE that we employ is a fairly simple one derived from the example VAE included in the Knet \cite{yuret2016knet} machine learning framework.
It is only slightly modified to run on images of generic size (the original VAE is hard-coded to run on $28\times28$ images from the MNIST \cite{deng2012mnist} data set).
We explored VAEs with $n_z=100$, $n_z=200$, and $n_z=400$.
The number of hidden neurons in each case was 500, 1000, and 2000, respectively.
We found that the $n_z=200$ VAE was about as well-trained as the $n_z=400$ VAE and significantly better trained than the $n_z=100$ VAE.
Similar behavior was observed when using these different VAEs to perform the inverse analysis.

\subsection{Inverse Analysis}
Our inverse analysis then amounts to combining the vartiational autoencoder and a forward model with an optimization problem formulated in terms of the latent variables, $\mathbf{z}$:
\begin{eqnarray}
	f(\mathbf{z}) &=& [h(d(\mathbf{z})) - \hat{\mathbf{h}}]^T \Sigma_h^{-1} [h(d(\mathbf{z})) - \hat{\mathbf{h}}] + [\mathbf{z}-\bar{\mathbf{z}}]^T\Sigma_z^{-1}[\mathbf{z}-\bar{\mathbf{z}}]
	\label{eq:ofcomplex} \\
	\hat{\mathbf{z}} &=& \argmin_{\mathbf{z}}~ f(\mathbf{z})
	\label{eq:argmin}
\end{eqnarray}
where $\Sigma_h$ is the covariance matrix of the observations, $\Sigma_z$ is the covariance matrix for the latent variables, $\bar{\mathbf{z}}$ is the mean of the latent variables, and we call $f(\mathbf{z})$ the objective function.
In many inverse analyses, it is assumed that $\Sigma_h$ is diagonal---effectively asserting that observational noises are uncorrelated.
We make this assumption and denote the $i^{\mathrm{th}}$ diagonal element of $\Sigma_h$ as $\sigma_{h,i}^2$.
The VAE seeks to make $\bar{\mathbf{z}}=0$ and $\Sigma_z=I$.
While it will have only ensured these equations are satisfied approximately, we assume that they are satisfied for the inverse analysis.
These assumptions simplify equation \ref{eq:ofcomplex} to
\begin{equation}
	f(\mathbf{z}) = \sum_{i=1}^{n_h}\frac{[h_i(d(\mathbf{z})) - \hat{h}_i]^2}{\sigma_{h,i}^{2}} + \sum_{i=1}^{n_z}z_i^2
	\label{eq:ofsimple}
\end{equation}
After an estimate of the latent variables, $\hat{\mathbf{z}}$, is obtained, the parameter field can then be estimated with the decoder,
\begin{equation}
	\hat{\mathbf{p}} = d(\hat{\mathbf{z}})
\end{equation}

We utilize finite differencing to compute $\nabla f(\mathbf{z})$.
Using forward or backward finite differencing makes the computational cost of computing $\nabla f(\mathbf{z})$ equivalent to $n_z+1$ evaluations of $f(\mathbf{z})$.
The dominant cost of each evaluation of $f(\mathbf{z})$ is typically the cost of computing $h(\cdot)$, i.e., it is dominated by the cost of the forward physical model.
Note that these evaluations can be performed in a perfectly parallel fashion without communication between the workers, making it well-suited to cloud computing.
Without some form of dimensionality reduction such as the VAE used here, these finite difference techniques are typically not applicable to highly parameterized models where $n_p$ is very large -- because $n_p+1$ evaluations of $f(\mathbf{p})$ would be required to compute $\nabla f(\mathbf{p})$.

Using the simplified objective function from equation \ref{eq:ofsimple} makes the regularization process straightforward even for models that are highly parameterized (i.e., $n_p$ is very large).
Since the latent variables approximately follow a $N(0, I)$ distribution, the natural regularization is the square of the Euclidean norm of $\mathbf{z}$.
The difficulties that are typically encountered in the formulation and computation of the regularization component of the objective function were essentially transferred to the data generation and VAE training components of our approach.
Our results show that a simple ``off-the-shelf'' VAE can be used effectively and trained easily.
Therefore, the difficulty associated with the regularization has been reduced to effective data generation.

Our perspective is that the data generation step is a good place to put the difficulty.
This is because it gives modelers a great deal of flexibility to describe their conceptual model of the structure of the parameters without having to concern themselves with both formulating a regularization penalty that conforms to this conceptual model and devising a way to efficiently compute this regularization penalty and its gradient.
This can be contrasted with, e.g., the geostatistical approach where clever computational methods must be utilized to efficiently compute the regularization component of the objective function for large scale problems.
Further, the creativity in formulating the conceptual model in the geostatistical approach is limited to formulating a variogram.
Here, the conceptual model is embodied in a computer program that generates unconditioned realizations of the parameter field, which is a very flexible approach.

With the ability to efficiently compute the gradient of $f(\mathbf{z})$, existing methods for gradient-based optimization can be used.
We have utilized the limited-memory Broyden-Fletcher-Goldfarb-Shanno \cite{liu1989limited} (L-BFGS) method with a Hager-Zhang line search \cite{hager2005new} for this purpose.
The version we use is implemented in the Optim.jl \cite{mogensen2018optim} software package for the Julia programming language.
In all cases, we begin the inverse analysis with an initial guess of $\mathbf{z}_0=\mathbf{0}$.

\section{Results}\label{sec:results}
\begin{figure}
	\includegraphics[width=\textwidth]{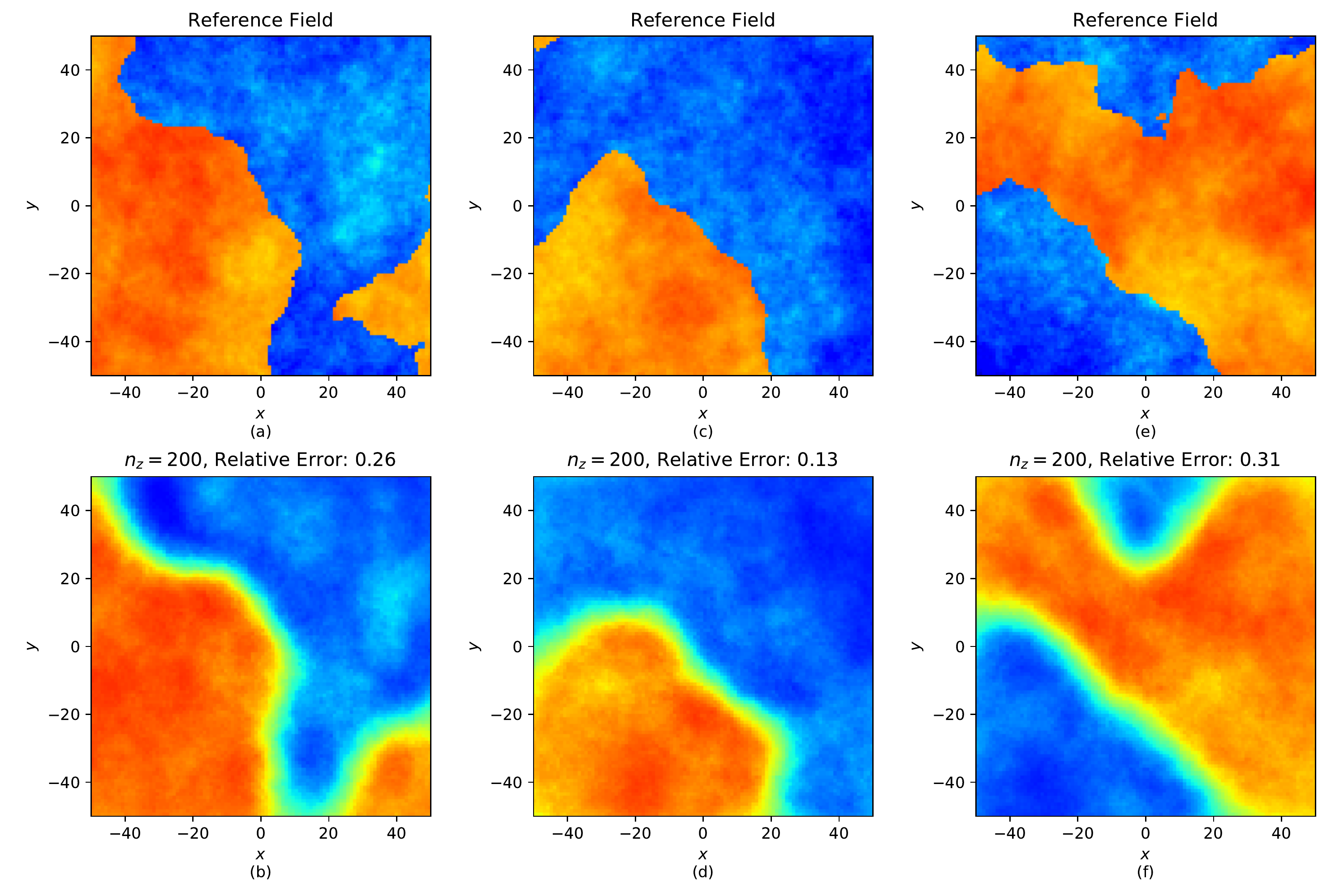}
	\caption{Three reference conductivity fields are shown in subfigures a, c, e and their corresponding inverse results using RegAE are in subfigures b, d, f.}
	\label{fig:comparison}
\end{figure}

We applied the methods described in the previous section to estimate the hydraulic conductivity of three synthetic aquifers.
Each of these reference ``true'' hydraulic conductivity fields (which are depicted in figure \ref{fig:comparison}a,c,e) was generated in the same manner that was used to generate the training data.
Of course, these reference conductivity fields where not part of the training set that was used to train the VAE.
However, they are generated using the same conceptual model of aquifer heterogeneity.
The observations that are used to inform the inverse analysis consist of observations of the hydraulic head and hydraulic conductivity on a $5\times5$ regular grid within the domain.
The forward model includes a steady-state flow simulation that produces predictions of the hydraulic head given a hydraulic conductivity input.
The hydraulic head observations were obtained using the model with the reference field as the input hydraulic conductivity.
The hydraulic conductivity observations were obtained directly from the reference field.
The forward model predictions for the hydraulic conductivity are obtained directly from the decoder.

The inverse result for the case with 200 latent variables (i.e., $n_z=200$) is shown in figure \ref{fig:comparison}b,d,f.
We use the relative error as a measure of how close the inverse result is to the reference field, and define the relative error as
\begin{equation*}
	\frac{||\mathbf{p}_r - d(\hat{\mathbf{z}})||^2}{||\mathbf{p}_r - \bar{\mathbf{p}}_r||^2}
\end{equation*}
where $\mathbf{p}_r$ is the reference field and $\bar{\mathbf{p}}_r$ is the mean of the reference field.
The relative error in the inverse analysis for each reference field with $n_z=200$ is 0.21, 0.14, and 0.31.
Inverse analyses were also performed for the cases of $n_z=100$ and $n_z=400$.
The relative error in the inverse analysis for each reference field when $n_z=100$ was 0.33, 0.18, and 0.38.
This indicates that there is a somewhat significant improvement in the quality of the inverse result increasing from $n_z=100$ to $n_z=200$.
By contrast, there was generally no improvement (and sometimes the relative error is higher) in increasing from $n_z=200$ to $n_z=400$.
The relative error in the inverse analysis for each reference field when $n_z=400$ was 0.22, 0.12, and 0.32.

\begin{figure}
	\includegraphics[width=\textwidth]{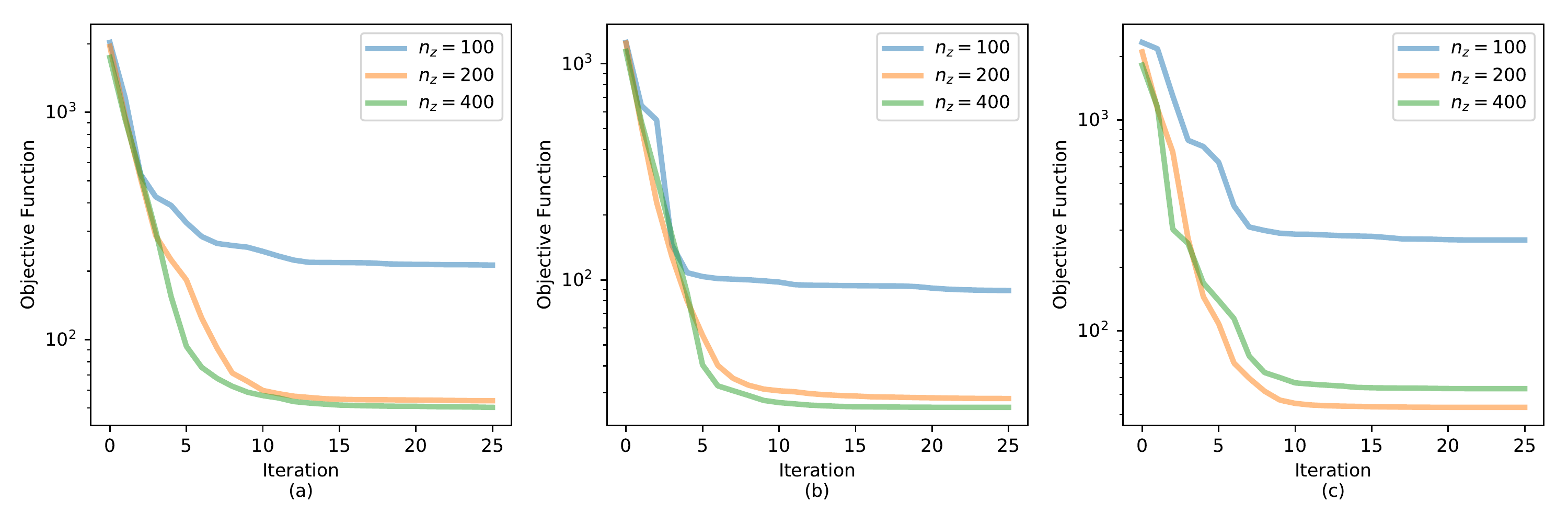}
	\caption{The convergence of the inverse analysis is shown for different values of $n_z$ for each of the three inverse analyses performed in figure \ref{fig:comparison}.}
	\label{fig:convergence}
\end{figure}

Figure \ref{fig:convergence} shows the convergence results for each of the $n_z=100$, $n_z=200$, and $n_z=400$ cases.
From this figure, one can see that convergence is generally obtained after $\sim$10 iterations with only minor improvements in the objective function after that.
The convergence plots also illustrate the trend of improvement in the objective function increasing from $n_z=100$ to $n_z=200$ but little or no improvement increasing from $n_z=200$ to $n_z=400$.
This is similar to the behavior of other methods that reduce the dimension of the parameter \cite{kitanidis2014principal,lee2014large,lin2017large} where there is little or no benefit in increasing the dimensionality of the reduced parameter space beyond a certain point.

Qualitatively, the inverse results for each of these examples captures the large-scale trend
That is, it captures where the ``Conductivity 1'' (orange) and ``Conductivity 2'' (blue) fields are present.
It also captures some of the character of the variability within each of these two fields.
However, the edges of the inverse results are blurred in comparison to the sharp edges between the orange and blue regions of the reference fields.

It is worth comparing results in figure \ref{fig:comparison} with figures 1 and 2 from previous work \cite{barajas2015linear} which explored inverse analysis using essentially the same observation network and a similar conceptual model of the hydraulic conductivity with high and low regions.
In the previous work, the inverse result is approximately a piecewise constant function with little variability within each of these pieces.
It smooths out small-scale variability, which is typical of most approaches to regularization, but the RegAE approach retains this small-scale variability.
However, the edges in the previous work are arguably sharper (unless one interprets the light blue region from figure 2 in the previous work as a large blurred edge).
RegAE does a better job of capturing the hydraulic conductivity within the low hydraulic conductivity regions, which are generally too high in figure 2 from the previous work, except at the grid points where the hydraulic conductivity is observed directly.
Overall the regularization from RegAE is at least comparable in performance and arguably better than the hand-crafted regularization that was constructed to solve the inverse problem in the previous work.
RegAE has several distinct advantages as well, namely parameter reduction (which eliminates the need for adjoint methods), the regularization does not have to be hand-crafted for a specific problem, and it does not introduce tuning parameters (such as the $\alpha$, $\beta$, $\gamma$, and $\delta$ of \cite{barajas2015linear}) into the inverse analysis.

Our analysis was performed on a machine with two 2.1GHz Intel Xeon 4116 CPUs with a total of 24 physical cores as well as an NVIDIA Quadro P5000 GPU.
The data generation in the case of $N=10^5$ realizations (which was used for the inverse analysis) required $\sim$186 seconds of wall time using 24 cores.
The GPU was used to train the VAE, and this process took $\sim$90 seconds, 144 seconds, and 250 seconds for the $n_z=100$, $n_z=200$, and $n_z=400$ cases, respectively.
The time to perform the inverse analysis varied somewhat depending on which reference field was used.
Averaging the time required to perform the inverse analysis on each of the three reference fields shown in figure \ref{fig:comparison}, it required $\sim$107 seconds, $\sim$196 seconds, and $\sim$305 seconds in the $n_z=100$, $n_z=200$, and $n_z=400$ cases, respectively.
Note that in the case of more complex models (e.g., transient flow, multiphase flow, reactive transport, etc), the cost of performing the inverse analysis will increase significantly because the cost of running the forward model will increase significantly.
However, the computational cost of the data generation and training the VAE should remain about the same, all else being equal.
Therefore, the use of RegAE generally does not add a significant computational cost to the inverse analysis.

\section{Discussion}\label{sec:discussion}
The only software specific to the individual problem being solved was the code for generating the unconditioned realizations of the parameters, $\mathbf{p}$, and the forward model, $h(\mathbf{p})$.
In the context of subsurface flow and transport, there are many excellent methods for generating unconditioned realizations of parameter fields and forward modeling.
The RegAE approach provides a means to assemble these pieces to effectively perform inverse analysis.
Our goal here is to demonstrate the basic concept of this approach as simply as possible.
We now focus our discussion on modifications of this approach that offer the potential for significant improvements.

The computational cost of the inverse analysis is dominated by the gradient calculations (especially when more expensive physical models are used), and this cost is proportional to $n_z$.
Because of this, the computational cost of performing the inverse analysis with $n_z=400$ is significantly higher than the cost of performing the inverse analysis with $n_z=200$.
However, it provides little or no improvement in the quality of the inverse analysis as measured by either the relative error or the objective function value.
Therefore, it is prudent to choose a value of $n_z$ that is large enough to obtain a good inverse result, but small enough to keep the computational cost down.
It is also possible to train a series of VAEs with increasing values of $n_z$, and to use larger values of $n_z$ as the inverse analysis proceeds.
For example, using the $n_z=100$ and $n_z=200$ VAEs here, one might begin by performing the first few iterations of the inverse analysis using the $n_z=100$ decoder.
Then, decode the result from that to obtain an estimate of $\mathbf{p}$, and run that estimate of $\mathbf{p}$ through the $n_z=200$ encoder.
The output of that encoding process could then be used to resume the inverse analysis using the $n_z=200$ decoder.

We also note that, while the dimensionality reduction here makes it possible to perform inverse analysis using highly parameterized models without adjoint methods, it does not preclude the possibility of using adjoint methods.
The reduction to $n_z$ parameters from $n_p$ calibration parameters makes this possible, but the cost of computing a gradient here is $\sim$200 model runs.
The cost of computing a gradient with adjoint methods is $\sim$2 model runs.
Therefore, if adjoint methods are available they could be used to further speed up these computations by an additional factor of up to $\sim100$.
When adjoint methods are available, the primary benefit of the RegAE approach will simply be to have an efficient form of regularization that is learned from unconditioned realizations of the parameter field.
In many cases sophisticated modeling codes such as FEHM \cite{zyvoloski1988fehm} or PFLOTRAN \cite{lichtner2015pflotran} have been developed to solve highly nonlinear sets of equations for physics such as multiphase flow and reactive transport, and it is difficult to retrofit these codes to exploit adjoint methods.
In such cases, the parameter reduction is a major advantage of the RegAE approach.

While a VAE could be regarded as a complex machine learning approach, we have deliberately attempted to use a simple VAE.
The neural networks that make up the encoder and decoder are shallow, non-convolutional, and fairly small.
Adding neurons and layers of neurons has the potential to improve the regularization and/or dimensionality reduction, but may require additional training data.
It may also require a more complex conceptual model of the structure of the parameters to justify its use.
Convolutional neural networks have proven to be powerful tools for a wide array of image processing tasks and we anticipate they could be powerful here as well.
One of the limitations of using a convolutional network in this context is that it would limit the applicability of the approach to problems on regular grids.
While our results are applied to a regular grid, the network we are using would also work for a problem with an unstructured grid.
Using a VAE that uses a perceptual loss function during the training process also has the potential to improve upon the results presented here.
In particular, using a perceptual loss function with a VAE has been shown to reduce blur in other image processing contexts \cite{hou2017deep}, and could potentially sharpen the edges between the high and low hydraulic conductivity regions in figure \ref{fig:comparison}b,d,f.

\section{Conclusion}\label{sec:conclusion}
We have presented an approach, called RegAE, for performing inverse analysis.
The main goal of this approach is to provide a means of regularizing inverse problems where the parameter fields are high-dimensional and have coherent structures.
RegAE leverages a VAE to learn how to regularize these problems based on unconditioned realizations of the high-dimensional parameter fields.
We demonstrated the approach for a hydrogeologic inverse problem where the structure is defined by two faces with distinct hydraulic conductivity distributions.
This approach provided a computationally efficient means of obtaining a good solution to this inverse problem, because, in addition to easing the regularization process, RegAE also reduces the dimensionality of the parameter space.
While the results here are encouraging, there remain many avenues to improve the performance of RegAE that we leave for future work.

\section*{Data Availability}
All the data used in this manuscript was automatically generated by a computer program.
The code for generating the data, training the VAE, and performing the inverse analysis is available at https://github.com/madsjulia/RegAE.jl

\bibliographystyle{unsrt}
\bibliography{refs}

\end{document}